\pdfoutput=1
\documentclass[trackchanges, twocolumn]{aastex7}

\usepackage[caption=false]{subfig}
\usepackage{graphicx}
\usepackage{bm}
\usepackage{amssymb,amsmath}
\usepackage{mathrsfs}
\usepackage{latexsym}
\usepackage{color}
\usepackage[normalem]{ulem}
\usepackage{dcolumn}
\usepackage[utf8]{inputenc}
\usepackage{microtype}
\usepackage{etoolbox}
\usepackage{accents}
\usepackage{multirow}
\usepackage{orcidlink}
\usepackage{makecell,tabularx,colortbl}
\usepackage{booktabs,cellspace}
\usepackage{multirow}
\usepackage{float}
\usepackage{xspace}

\allowdisplaybreaks

\definecolor{dodgerblue}{HTML}{1E90FF}
\definecolor{viennared}{HTML}{DA0A14}
\definecolor{ctorange}{HTML}{FF6C0C}
\definecolor{wales}{HTML}{ff0038}
\definecolor{benettongreen}{HTML}{009421}
\definecolor{ferrarired}{HTML}{ff2800}
\definecolor{austriawienpurple}{HTML}{441678}
\definecolor{gray}{HTML}{F0F0F0}
\definecolor{LightCyan}{rgb}{0.88,1,1}
\definecolor{headerblue}{RGB}{0,112,192}
\definecolor{lightgray}{gray}{0.9}
\definecolor{ZurichBlue}{rgb}{.255,.41,.884} 		% RoyalBlue of svgnames
\definecolor{ZurichRed}{rgb}{0.9, 0.1, 0} 			% Red of svgnames
\definecolor{ZurichGreen}{rgb}{.196,.504,.396} 		% LimeGreen of svgnames
\definecolor{ZurichYellow}{rgb}{1,.648,0} 			% Orange of svgnames
\definecolor{dodgerblue}{rgb}{0.12, 0.56, 1.0}
\definecolor{azure}{rgb}{0.0, 0.5, 1.0}
\definecolor{alizarincrimson}{rgb}{0.82, 0.1, 0.26}
\definecolor{mediumpurple}{rgb}{0.58, 0.44, 0.86}
\definecolor{lasallegreen}{rgb}{0.03, 0.47, 0.19}
\definecolor{my_gray}{rgb}{0,0,0}
\definecolor{uniwienblue}{HTML}{006699}
\definecolor{walesred}{HTML}{ff0038}
\definecolor{myorange}{HTML}{FF6C0C}
\definecolor{ferngreen}{HTML}{009246}
\definecolor{scarletred}{HTML}{CD212A}

\newcolumntype{a}{>{\columncolor{gray}}c}
\newcolumntype{b}{>{\columncolor{white}}c}

\newcommand{\rateALNSBH}{104.3^{+118.3}_{-61.8}}
\newcommand{\rateQCNSBH}{92.2^{+123.4}_{-59.6}}

\newcommand{\massprimarymed}{11.5}
\newcommand{\massprimaryupper}{0.6}
\newcommand{\massprimarylower}{1.6}
\newcommand{\massprimary}{\massprimarymed^{+\massprimaryupper}_{-\massprimarylower}}

\newcommand{\chiprecmed}{0.06}
\newcommand{\chiprecupper}{0.13}
\newcommand{\chipreclower}{0.04}
\newcommand{\chiprec}{\chiprecmed^{+\chiprecupper}_{-\chipreclower}}

\newcommand{\chieffecmed}{0.005}
\newcommand{\chieffecupper}{0.071}
\newcommand{\chieffeclower}{0.095}
\newcommand{\chieffec}{\chieffecmed^{+\chieffecupper}_{-\chieffeclower}}

\newcommand{\spinprimarymed}{0.06}
\newcommand{\spinprimaryupper}{0.18}
\newcommand{\spinprimarylower}{0.05}
\newcommand{\spinprimary}{\spinprimarymed^{+\spinprimaryupper}_{-\spinprimarylower}}

\newcommand{\setdata}{\lbrace d \rbrace}

\newcommand{\beq}{\begin{equation}}
\newcommand{\eeq}{\end{equation}}

\newcommand{\pyEFPE}{\textsc{pyEFPE}\xspace}
\newcommand{\XP}{\textsc{IMRPhenomXP}\xspace}
\newcommand{\XPHM}{\textsc{IMRPhenomXPHM}\xspace}

\newcommand{\uam}{Instituto de F\'isica Te\'orica UAM/CSIC, Universidad Aut\'onoma de Madrid, Cantoblanco 28049 Madrid, Spain}
\newcommand{\bham}{School of Physics and Astronomy and Institute for Gravitational Wave Astronomy, University of Birmingham, Edgbaston, Birmingham, B15 2TT, United Kingdom}
\newcommand{\AEI}{Max Planck Institute for Gravitational Physics (Albert Einstein Institute), D-14467 Potsdam, Germany}

\begin{document}

%---------- Titles
\title{Impact of eccentricity on the population properties of neutron star -- black hole mergers}

%---------- Authors
\author[orcid=0000-0002-9977-8546]{Gonzalo Morras}
\affiliation{\AEI}
\affiliation{\uam}
\email[]{gonzalo.morras@uam.es}

\author[orcid=0000-0003-4984-0775]{Geraint Pratten}
\affiliation{\bham}
\email[show]{g.pratten@bham.ac.uk}

\author[0000-0003-1542-1791]{Patricia Schmidt}
\affiliation{\bham}
\email[]{p.schmidt@bham.ac.uk}

%---------- Abstract
\begin{abstract}
We revisit the population properties of neutron star-black hole (NSBH) mergers using low-mass compact binary coalescences reported through GWTC‑4. 
Employing \pyEFPE, an inspiral-only waveform model that captures both orbital eccentricity and spin-induced precession, we reanalyse all binary neutron star (BNS) and NSBH events observed via gravitational waves. 
The BNS systems GW170817 and GW190425 are fully consistent with quasi-circular inspirals, while GW200105 stands out among the NSBH binaries as the only system exhibiting significant residual eccentricity at $20\,\mathrm{Hz}$, strengthening evidence for dynamically driven formation pathways. 
The remaining NSBH events show no measurable eccentricity and appear broadly compatible with low-spin binaries formed through isolated stellar evolution.
Using hierarchical Bayesian inference, we obtain the first joint constraints on the mass, spin, and eccentricity distributions of NSBH binaries.
Our results also yield the first simultaneous constraints on spin precession and orbital eccentricity in NSBH mergers, while the inferred merger rates remain fully consistent with previous LVK measurements.
Treating all NSBH systems as a single population yields results compatible with formation in hierarchical triples, whereas the quasi-circular population remains broadly consistent with isolated evolution.
Our results highlight the emerging role of eccentricity as a key discriminator between formation channels. As the number of NSBH detections grows, joint constraints on masses, spins, and orbital eccentricity will enable increasingly sharp tests of dynamical versus isolated binary evolution, establishing NSBH systems as powerful probes of compact-object astrophysics.
\end{abstract}


%---------- Main Article

%---------- Introduction
\section{Introduction}
Gravitational-wave (GW) observations with the Laser Interferometer Gravitational-Wave
Observatory (LIGO)~\citep{LIGOScientific:2014pky}, Virgo~\citep{VIRGO:2014yos} and KAGRA~\citep{KAGRA:2020tym} (LVK) are a unique window into astrophysical populations of compact objects, allowing us to characterize the underlying distributions of their masses, spins, eccentricities, and their redshift-dependent merger rates. 
These population properties encode the physics of stellar evolution, binary dynamics, and compact object formation. 
Identifying the binary formation channels is one of the largest open questions in modern astrophysics, with important consequences for stellar physics, binary evolution, and relativistic astrophysics. 

The groundbreaking observation of the mergers between neutron stars and black holes~\citep{LIGOScientific:2021qlt} provided unambiguous evidence for the existence of these systems, which has been strengthened by subsequent observations~\citep{KAGRA:2021vkt,LIGOScientific:2024elc,LIGOScientific:2025slb}. 
This has triggered a series of investigations into the astrophysical properties of neutron star-black hole (NSBH) mergers and neutron star physics~\citep{Farah:2021qom, Landry:2021hvl, Ye:2022qoe, Zhu:2021jbw, Biscoveanu:2022iue}. It is of particular interest to disentangle the different scenarios for the evolutionary pathways of NSBH mergers. 

Although numerous models have been proposed, see~\cite{Broekgaarden:2021iew} and references therein, the exact astrophysical processes at work remain highly uncertain.
In the currently favored canonical scenario, NSBH systems form through isolated evolution of massive binaries undergoing a common envelope phase~\citep{Belczynski:2001uc,Mapelli:2018wys,Drozda:2020kqz,Broekgaarden:2021iew,Broekgaarden:2021hlu,Mandel:2021ewy,Xing:2023zfo}.
This scenario leads to several core predictions for the astrophysical properties of these binaries.
For example, natal black holes may be born with low spins if angular momentum transport between the helium core and hydrogen envelope, combined with envelope stripping, efficiently removes angular momentum from the progenitor~\citep{Qin:2018vaa,Fuller:2019sxi}.
While black holes are expected to form at wide orbits, tidal interactions in close binaries can potentially induce significant spin~\citep{van2007long,Kushnir:2016zee,Qin:2018vaa,Mandel:2020lhv}.
The resulting spins are expected to be approximately aligned with the orbital angular momentum due to tidal interactions, though supernova kicks can induce misalignment~\citep{Kalogera:1999tq,Wysocki:2017isg,Giacobbo:2019fmo,Oh:2023hgh}. 
These binaries should have negligible orbital eccentricity by the time they enter the LIGO-Virgo-KAGRA (LVK) sensitivity band, as eccentricity is very efficiently radiated away~\cite{Peters:1964zz}.
However, it is important to note that mechanisms such as mass transfer~\citep{Rocha:2024oqc,Zenati:2024elj} and supernova kicks~\citep{Brandt:1995snk,Kalogera:1996rm,Vigna-Gomez:2025vjk} can challenge the assumption that residual eccentricity must always be negligible in isolated binary evolution.

In contrast, evolutionary pathways that involve dynamical interactions can lead to markedly different predictions for the properties of NSBH binaries. 
A wide range of such dynamical channels have been proposed.
First, NSBH systems may form through dynamical assembly in dense stellar environments, including globular clusters \citep{PortegiesZwart:1999nm,Freire:2004sr,Ye:2019luh,Zevin:2018kzq,Sedda:2020wzl,Fragione:2020wac,Dorozsmai:2025jlu} and young stellar clusters \citep{PortegiesZwart:1999nm,Rastello:2020sru,Trani:2021tan}, where repeated few‑body encounters can assemble compact binaries or form hierarchical triples through dynamical interactions.
In globular clusters, however, the NSBH merger rate is expected to be highly suppressed.
Ultra‑wide binaries may be driven to merger through secular torques from the Galactic potential \citep{Michaely:2022ujf,Stegmann:2024rnk}, though the resulting rates depend sensitively on assumptions about natal kicks \citep{Michaely:2022ujf}. 
Additional dynamical mechanisms include fly‑by interactions and direct gravitational‑wave captures \citep{Hoang:2020gsi}, although these channels generally predict comparatively low merger rates.
Active galactic nuclei offer another pathway, in which gas torques and migration traps can efficiently drive compact binaries toward coalescence \citep{Stone:2016wzz,Yang:2020xyi,McKernan:2020lgr}, but these predictions come with substantial astrophysical uncertainties.
Finally, NSBH mergers can arise in hierarchical multiple systems, particularly stellar triples \citep{Silsbee:2016djf,Antonini:2017ash,Hamers:2019oeq,Fragione:2019zhm,Trani:2021tan,Stegmann:2025clo}, some of which naturally form in young clusters (hence the dual appearance of \citealt{Trani:2021tan}). 
In such systems, a bound tertiary companion induces secular gravitational perturbations through the von Zeipel–Kozai–Lidov (ZKL) mechanism \citep{Zeipel:1910,Lidov:1962,Kozai:1962}, generating large‑amplitude eccentricity oscillations that can both accelerate the inspiral and significantly enhance the merger rate, while imprinting distinctive signatures such as residual eccentricity and spin–orbit misalignment. 
This final class is particularly appealing given the observational support that most massive progenitor stars are in hierarchical triples~\citep{Moe:2017str}.

A key feature of dynamical formation channels is that they can naturally predict substantial spin-orbit misalignment and, in many cases, non-negligible residual orbital eccentricity even at relatively high gravitational-wave frequencies. 
The recent observation of measurable eccentricity in GW200105~\citep{Morras:2025xfu}, one of the NSBH binaries reported by the LVK, is therefore difficult to reconcile with expectations from isolated binary evolution. 
Any eccentricity observed in the detector band must have been even larger at earlier times due to the rapid circularization driven by gravitational-wave emission~\citep{Peters:1964zz}. 
This strengthens the case for exploring alternative, dynamically driven formation scenarios.

Recent work has strengthened this tension. 
In particular, \cite{Stegmann:2025clo} show that the combination of residual eccentricity, spin-orbit misalignment, and the inferred NSBH merger rate is naturally explained
if a significant fraction of NSBH mergers originate from hierarchical triple systems.
The three-body dynamics of compact unequal-mass triples drives mergers across a broad range of orbital parameters whilst avoiding finely tuned tertiary inclinations.
Such systems can retain measurable eccentricity down to gravitational-wave frequencies, providing a compelling formation pathway for events like GW200105.

Simultaneously, tentative hints for eccentricity are emerging in the binary black hole (BBH) population~\citep{Romero-Shaw:2022xko,Gupte:2024jfe,Planas:2025jny,Xu:2025ajj}, though recent analyses also suggest that the branching ratio for eccentric events is below $0.0239$ at $90\%$ confidence~\citep{Zeeshan:2026pga}, with the rate of eccentric binaries being only weakly constrained by observations and remaining highly model-dependent. 

Motivated by these developments, we revisit the population properties of NSBH binaries using gravitational-wave data alone. 
In particular, we reanalyse all low-mass events observed to date to assess whether current observations already favour a multi-channel formation picture or remain consistent with a predominantly isolated binary origin. 
We perform a joint analysis of the mass, spin, and eccentricity distributions, as this remains one of the most powerful ways to discriminate between the plethora of formation channels for compact binaries~\citep{Callister:2024cdx,Stegmann:2025shr,Stegmann:2025clo}.
An advantage to our analysis is that we use a recently developed waveform model, \pyEFPE~\citep{Morras:2025nlp}, that allows us to jointly infer spin-precession and orbital eccentricity.
Due to the extremely limited number of BNS observations to date, we only report individual parameter estimation results for these binaries and do not attempt to model the astrophysical population.

%---------- Analysis
\section{Analysis Methods}
\label{sec:Methods}

In this section, we describe the methodology used to analyze the GW events. This includes the parameter estimation framework, the waveform models employed, the selection criteria for the events included in this study, and the specific configuration used for their analyses.

\subsection{Parameter Estimation Framework}
\label{sec:Methods:PE}

We perform coherent Bayesian parameter estimation to infer the source properties of each GW event. Parameter inference is carried out using nested sampling~\citep{Skilling:2006gxv}, as implemented in \textsc{Dynesty}~\citep{dynesty} within the \textsc{Bilby} inference framework~\citep{bilby_paper}. This allows us to sample the posterior distribution
\begin{align}
    p(\bm{\lambda} \mid d) &=
    \frac{\mathcal{L}(d \mid \bm{\lambda})\, \pi(\bm{\lambda})}{\mathcal{Z}(d)},
\end{align}
where $\mathcal{L}(d \mid \bm{\lambda})$ is the likelihood of observing the data $d$ given source parameters $\bm{\lambda}$, $\pi(\bm{\lambda})$ the prior, and $\mathcal{Z}(d)$ the Bayesian evidence.

Assuming stationary Gaussian noise in each detector, the likelihood of observing a GW signal $h(\bm{\lambda})$ in a network of $N$ detectors is~\citep{Veitch:2014wba}
\begin{align}
    \mathcal{L}(d \mid \boldsymbol{\lambda})
    &\propto
        \exp\!\left[ -\frac{1}{2}
        \sum_{i=1}^{N}
        \langle h_i(\boldsymbol{\lambda}) - d_i
        \mid
        h_i(\boldsymbol{\lambda}) - d_i \rangle \right],
\end{align}
where $h_i(\bm{\lambda})$ denotes the predicted gravitational wave signal in the $i$-the detector and $d_i$ the corresponding strain data. 
The noise-weighted inner product $\langle \cdot \mid \cdot \rangle$ is defined as
\begin{equation}
    \langle a \mid b \rangle = 4 \, \Re \int_{f_{\rm low}}^{f_{\rm high}} \frac{\tilde{a}(f) \tilde{b}^*(f)}{S_n(f)} \, df,
    \label{eq:inner_prod}
\end{equation}
with $\tilde{a}(f)$ and $\tilde{b}(f)$ denoting the Fourier transforms of the time-domain functions $a(t)$ and $b(t)$, and $S_n(f)$ the one-sided power spectral density of the detector noise. 
The integration is performed over the frequency interval $[f_{\rm low}, f_{\rm high}]$.

\subsection{Waveform Models}
\label{sec:Methods:WF}

As described in Sec.~\ref{sec:Methods:PE}, Bayesian parameter estimation requires predictions of the GW signal $h(\bm{\lambda})$ as a function of the source parameters $\bm{\lambda}$. Since a single analysis involves evaluating the likelihood tens of millions of times, it is essential to employ waveform models that are not only accurate but also computationally efficient. However, the general relativistic two-body problem is very complex and rich in physical effects, and even the most advanced waveform models make approximations to remain tractable. Models that are fast enough for large-scale inference often neglect key features such as orbital eccentricity, spin-induced precession, higher-order harmonics, or the merger-ringdown phase. To capture the relevant physics across different scenarios, we use three different waveform models:

\begin{itemize}
    \item \pyEFPE~\citep{Morras:2025nlp}: A frequency-domain, post-Newtonian (PN) model for inspiralling precessing-eccentric compact binaries. This waveform omits the merger-ringdown phase and is valid only for frequencies below the innermost stable circular orbit (ISCO). 
    It models the GW amplitudes at Newtonian-order, corresponding to only having the $(l,|m|) = (2,2)$ and $(2,0)$ modes for eccentric binaries. 
    The model incorporates eccentric harmonics through closed-form Fourier-Bessel amplitudes.
    \item \XP~\citep{Pratten:2020ceb}: A frequency-domain model for the full inspiral-merger-ringdown of precessing, quasi-circular binaries. The inspiral is based on the PN formalism, while the merger and ringdown are described using a phenomenological ansatz. This waveform is calibrated against effective-one-body and numerical relativity simulations. In the co-precessing frame~\citep{Schmidt:2010it}, \XP includes only the leading $(l,|m|) = (2,2)$ mode~\citep{Pratten:2020fqn}.
    \item \XPHM~\citep{Pratten:2020ceb}:  An extension of \XP that includes higher-order modes (HMs)~\citep{Garcia-Quiros:2020qpx}. Specifically, in the co-precessing frame it includes the $(l,|m|) = (2,2)$, $(2,1)$, $(3,3)$, $(3,2)$, and $(4,4)$ modes.
\end{itemize}

We note that all models used in this work describe binary black holes. In particular, they neglect tidal deformability and other matter effects that could be relevant if one or both components were neutron stars.

\subsection{Event Selection}
\label{sec:Methods:Events}

In this work, we study orbital eccentricity and spin-induced precession in a subset of GW events observed by the LVK detector network. 
We focus on low-mass binary coalescences, where at least one component is consistent with a NS, as their signals are dominated by the inspiral phase, where these effects are both most detectable and can be accurately modeled using \pyEFPE. 
For such systems, the merger and ringdown occur at high frequencies, beyond the most sensitive band of current ground-based detectors, making them particularly well-suited for analysis with inspiral-only waveform models.

We restrict our analyses to events reported up to the GWTC-4 catalog~\citep{LIGOScientific:2018mvr,LIGOScientific:2020ibl,LIGOScientific:2021usb,KAGRA:2021vkt,LIGOScientific:2025slb}, which includes detections up to the first part of the fourth LVK observing run (O4a) and represents the most recent publicly available dataset at the time of writing.
Based on these considerations, we include in our study the two binary neutron star (BNS) mergers GW170817~\citep{LIGOScientific:2017vwq} and GW190425~\citep{LIGOScientific:2020aai}, as well as five neutron star–black hole (NSBH) mergers: GW190426\_152155, GW190917\_114630~\citep{LIGOScientific:2021usb}, GW200105\_162426, GW200115\_042309~\citep{LIGOScientific:2021qlt}, and GW230529\_181500~\citep{LIGOScientific:2024elc}. For brevity, we omit the timestamps after the underscores when referring to these events throughout the rest of the text. 

The selected events span total masses between $\approx 2.7\,M_\odot$ for GW170817~\citep{LIGOScientific:2018hze} and $\approx 12\,M_\odot$ for GW190917~\citep{LIGOScientific:2021usb}. Their signal-to-noise ratios (SNRs) range from $\approx 9.5$ for GW190917~\citep{LIGOScientific:2021usb} to $\approx 32.4$ for GW170817~\citep{LIGOScientific:2017vwq}.

We exclude some of the LVK-reported events with component masses consistent with containing at least one neutron star. In particular, GW191219\_163120 and GW200210\_092254~\citep{KAGRA:2021vkt} are excluded due to their significantly higher total masses ($\approx 32\,M_\odot$ and $\approx 27\,M_\odot$, respectively) and moderate SNRs ($\approx 8.9$ and $\approx 9.5$ respectively), which limits the applicability of inspiral-only models. GW230518\_125908~\citep{LIGOScientific:2025slb} is excluded because it was identified during the engineering run preceding O4a. Finally, we exclude GW190814~\citep{LIGOScientific:2020zkf}, which has a $2.59^{+0.08}_{-0.09}\,M_\odot$ secondary, slightly above the expected maximum neutron star mass~\citep{Rezzolla:2017aly,Margalit:2017dij,Shibata:2019ctb,Fan:2023spm}. Despite its relatively high total mass ($\approx 26\,M_\odot$), the event has a large total SNR ($\approx 25$), with a substantial amount of it accumulated during the inspiral. However, we do not include it in our analysis, as the strong contribution of higher-order modes, absent in \pyEFPE, could bias the inferred parameters. 

%%%%%%%%%%%%%%%%%%%%%%%%%%%%%%%%%%%%%%%%%%%%%%%%%%%
\subsection{Data Selection and Analysis Settings}
\label{sec:Methods:Settings}
%%%%%%%%%%%%%%%%%%%%%%%%%%%%%%%%%%%%%%%%%%%%%%%%%%%
We analyze the selected events using publicly available strain data, noise power spectral densities, and calibration envelopes~\citep{GW170817_LIGO_Virgo_2017,LIGOScientific:2019lzm,KAGRA:2023pio,LIGO_T1900685,LIGO_P2000026,LIGO_P2000223,ligo2021_gwtc2p1_PE_samples,ligo2021_gwtc3_glitchmodel,ligo2021_gwtc3_PE_samples,GW230529_data,LIGOScientific:2024elc}. For events affected by known noise transients, we use the corresponding BayesWave-deglitched strain data~\citep{Cornish:2014kda, Cornish:2020dwh}. In particular, GW170817 was affected by a prominent blip glitch in the LIGO-Livingston detector less than $1\,\mathrm{s}$ before merger. GW190425 had a short glitch in LIGO-Livingston about $60\,\mathrm{s}$ before merger. GW200105 exhibited minor scattered light noise artifacts in LIGO-Livingston below $25\,\mathrm{Hz}$, occurring approximately $3\,\mathrm{s}$ before merger. Finally, GW200115 was affected by more pronounced scattered light noise in LIGO-Livingston below $25\,\mathrm{Hz}$.

We set the low-frequency cut-off to $f_\mathrm{low} = 20\,\mathrm{Hz}$ for all events except for GW170817. For this event, we use a slightly higher value of $f_\mathrm{low} = 23\,\mathrm{Hz}$, both because it was observed during the second observing run (O2), when low-frequency sensitivity was significantly worse than in O3 and O4~\citep{aLIGO:2020wna,Capote:2024rmo}, and to limit the waveform duration to less than $128\,\mathrm{s}$ to reduce the computational cost consistent with the analyses presented by the LVK~\citep{LIGOScientific:2018mvr}.

The high-frequency cut-off, $f_\mathrm{high}$, is conservatively set below the frequency of the minimum energy circular orbit (MECO)~\citep{Cabero:2016ayq,Pratten:2020ceb} obtained from the posterior samples from publicly available analyses. This choice ensures that the analysis remains within the regime of validity of the PN approximations used in \pyEFPE. The value of $f_\mathrm{high}$ used for each event is listed in Table~\ref{table:AnalysisConfig}. For NSBH events, $f_\mathrm{high}$ is chosen just below $f_\mathrm{MECO}$, while for BNS systems (GW170817 and GW190425), we adopt an even more conservative value of $f_\mathrm{high} = 400\, \mathrm{Hz}$, well below $f_\mathrm{MECO}$, to avoid the high-frequency regime where tidal effects, absent in the waveform models employed, can play a significant role in nearly equal-mass systems~\citep{Flanagan:2007ix, Harry:2018hke,Pratten:2021pro}. 

We adopt broad, uninformative priors in the parameter estimation analyses. These priors are designed to cover the entire region of parameter space where the likelihood has significant support, helping avoiding biases in parameter recovery and ensuring the posteriors are suitable for reweighting in population analyses. Specifically, we use priors that are uniform in the detector-frame component masses and dimensionless spin magnitudes $\chi_1,\chi_2 \in [0, 0.8]$, isotropic in spin orientations, binary inclination, and sky position, and uniform in comoving volume and source-frame coalescence time. The component mass priors are constrained to ensure a mass ratio mass ratio $q = m_2/m_1 \in [0.05, 1]$ and a detector-frame chirp-mass that spans the range listed in Table~\ref{table:AnalysisConfig}. 
For analyses including orbital eccentricity, we adopt a uniform prior on the eccentricity $e_{20\mathrm{Hz}} \in [0, 0.4]$, defined at a reference GW frequency of 20 Hz, and a uniform prior on the mean anomaly. 

The duration of the data segment analyzed for each event, also shown in Table~\ref{table:AnalysisConfig}, is chosen to fully contain the longest possible signal given the prior bounds and frequency range. 

\begin{table}[]
    \centering
    \begin{tabular}{c|c|c|c}
        Event & $f_\mathrm{high}$ $[\mathrm{Hz}]$ & $\mathcal{M}_c^\mathrm{det}$ $[M_\odot]$ & Duration $[\mathrm{s}]$ \\ % & Data \\
        \hline
        GW170817 & $400$ & $[1.195 , 1.200]$ & $128$ \\ % &
        GW190425 & $400$ & $[1.475 , 1.495]$ & $128$ \\ % & 
        GW190426 & $350$ & $[2.4 , 2.8]$ & $64$ \\ % & 
        GW190917 & $190$ & $[3.6 , 4.6]$ & $32$ \\ % & 
        GW200105 & $340$ & $[3.45, 3.70]$ & $32$ \\ % & 
        GW200115 & $370$ & $[2.4 , 2.8]$ & $64$ \\ % & 
        GW230529 & $566$ & $[2.01 , 2.04]$ & $128$ \\ % & 
        \hline
    \end{tabular}
    \caption{High-frequency cut-off ($f_\mathrm{high}$) in $\mathrm{Hz}$, detector-frame chirp mass ($\mathcal{M}_c^\mathrm{det}$) prior range in solar masses ($M_\odot$), and segment duration in seconds (s) used in the parameter estimation analyses for each event.}
    \label{table:AnalysisConfig}
\end{table}

%---------- Results
\section{Results}
\label{sec:Results}

In this section, we present the results of the single-event eccentric-precessing parameter estimation analyses, separating the discussion of the BNS and NSBH systems. 
We report the following parameters: the source-frame component masses $m^{\rm source}_i$, the eccentricity $e_{20\rm Hz}$ defined at a GW reference frequency of $20\,\mathrm{Hz}$, the luminosity distance $D_L$ in $\mathrm{Mpc}$, and the two effective spin parameters that capture the dominant spin-orbit contributions to the gravitational-wave signal.
The first effective parameter is the mass-weighted projection of the component spins along the direction of the Newtonian orbital angular momentum, $\hat{L}_N$~\citep{Damour:2001tu,Racine:2008qv,Ajith:2009bn},
\begin{align}
\chi_{\rm eff} 
&= \left( \frac{m_1}{M} \boldsymbol{\chi}_1 
      + \frac{m_2}{M} \boldsymbol{\chi}_2 \right) 
      \cdot \hat{L}_N,
\end{align}
where $M = m_1 + m_2$ and $\boldsymbol{\chi}_i = \mathbf{S}_i / m_i^2$.
The second is an effective precession parameter defined at the reference frequency, which quantifies the degree of relativistic precession induced by spin-orbit misalignment~\citep{Schmidt:2014iyl},
\begin{align}
\chi_p 
&= \max \left[
    \chi_1 \sin \theta_1,\,
    \left( \frac{3 + 4q}{4 + 3q} \right)
    q \, \chi_2 \sin \theta_2
    \right],
\end{align}
where $q = m_2/m_1 \le 1$ and $\theta_i$ denotes the tilt angle between $\boldsymbol{\chi}_i$ and $\hat{L}_N$.

\subsection{Binary Neutron Stars}
\label{sec:Results:BNS}

\begin{figure*}
    \centering
    \includegraphics[width=1\textwidth]{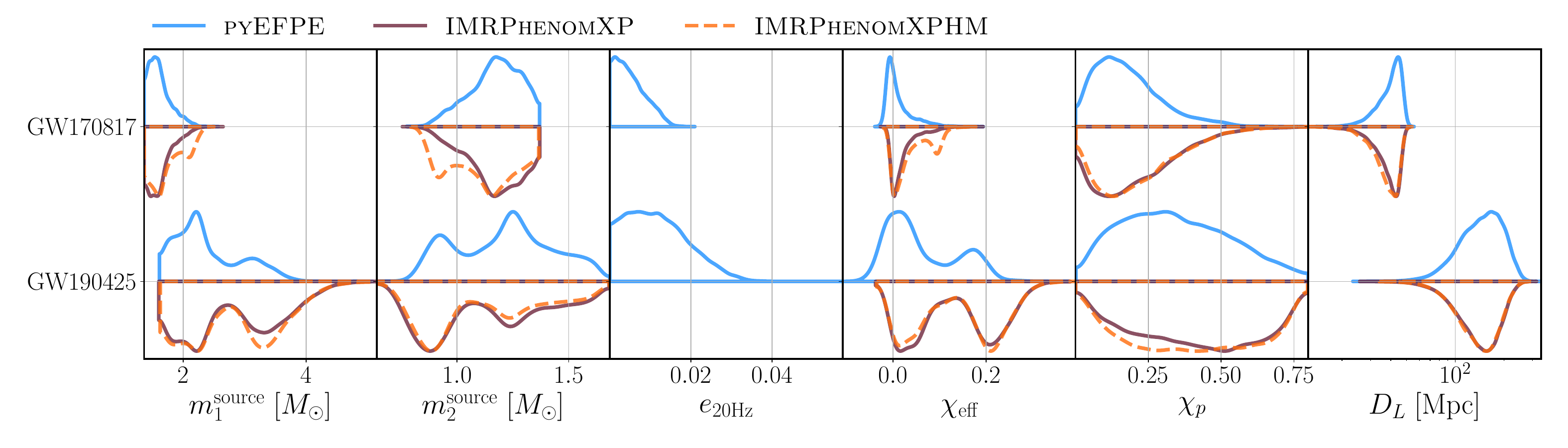}
    \caption{Marginal posterior distributions for selected source parameters of the two BNS events. Shown are the source-frame component masses ($m_1^\mathrm{source}$ and $m_2^\mathrm{source}$), eccentricity at 20Hz ($e_{20\mathrm{Hz}}$), the effective inspiral spin ($\chi_\mathrm{eff}$), the effective precession spin ($\chi_p$), and the luminosity distance ($D_L$). In each violin plot, the upper half corresponds to results from the eccentric \pyEFPE analysis, while the lower half shows posteriors from the quasi-circular \XP (solid) and \XPHM (dashed) analyses.}
    \label{fig:BNSs_violins}
\end{figure*}

The marginal posterior distributions for selected source parameters of the BNS events are shown in Fig.~\ref{fig:BNSs_violins}. Due to their long inspiral durations in band, the parameters of these systems are tightly constrained, particularly for GW170817, which was observed with an SNR of approximately 32~\citep{LIGOScientific:2017vwq}. We find that both BNS events are consistent with a quasi-circular inspiral. At the 90\% credible level, we constrain the eccentricities at 20\,Hz to $e_{20\mathrm{Hz}}^\mathrm{GW170817} < 0.011$ and $e_{20\mathrm{Hz}}^\mathrm{GW190425} < 0.022$, respectively. We report eccentricity constraints at 20\,Hz because this is the frequency from which the data are analyzed, thereby avoiding extrapolation to unobserved frequencies.

Nonetheless, for comparison with the literature, we can convert these constraints to a reference frequency of 10\,Hz using that~\citep{Peters:1964zz}
\begin{equation}
    \frac{f_1}{f_2} = \left( \frac{e_2}{e_1} \right)^{18/19} \left( \frac{1 - e_2^2}{1 - e_1^2} \right)^{-3/2} \left(\frac{1 + \frac{121}{304} e_2^2}{1 + \frac{121}{304} e_1^2} \right)^{1305/2299} \, ,
\end{equation}
\noindent where $e_1$ and $e_2$ correspond to the eccentricities measured at frequencies $f_1$ and $f_2$, respectively. Using this relation, the 90\% credible upper limits translate to $e_{10\mathrm{Hz}}^\mathrm{GW170817} < 0.022$ and $e_{10\mathrm{Hz}}^\mathrm{GW190425} < 0.046$, in agreement with previous constraints reported in~\citet{Lenon:2020oza}.

Consistent with the lack of evidence for eccentricity, Fig.~\ref{fig:BNSs_violins} shows that the parameters estimated with \pyEFPE are in good agreement with those obtained using the \XP and \XPHM approximants. 
In particular, GW170817 has component masses compatible with neutron stars observed in Galactic binary systems~\citep{Tauris:2017omb}, and spins that are consistent with being small. 
In contrast, GW190425 favors larger component masses which, while still compatible with neutron stars, are not representative of the observed Galactic binary neutron star population~\citep{Landry:2021hvl}. 
Moreover, due to its significantly lower SNR, the spin parameters are only weakly constrained.

%%%%%%%%%%%%%%%%%%%%%%%%%%%%%%%%%%%%%%%%%%
\subsection{Neutron Star - Black Holes}
\label{sec:Results:NSBH}
%%%%%%%%%%%%%%%%%%%%%%%%%%%%%%%%%%%%%%%%%%
\begin{figure*}
    \centering
    \includegraphics[width=1\textwidth]{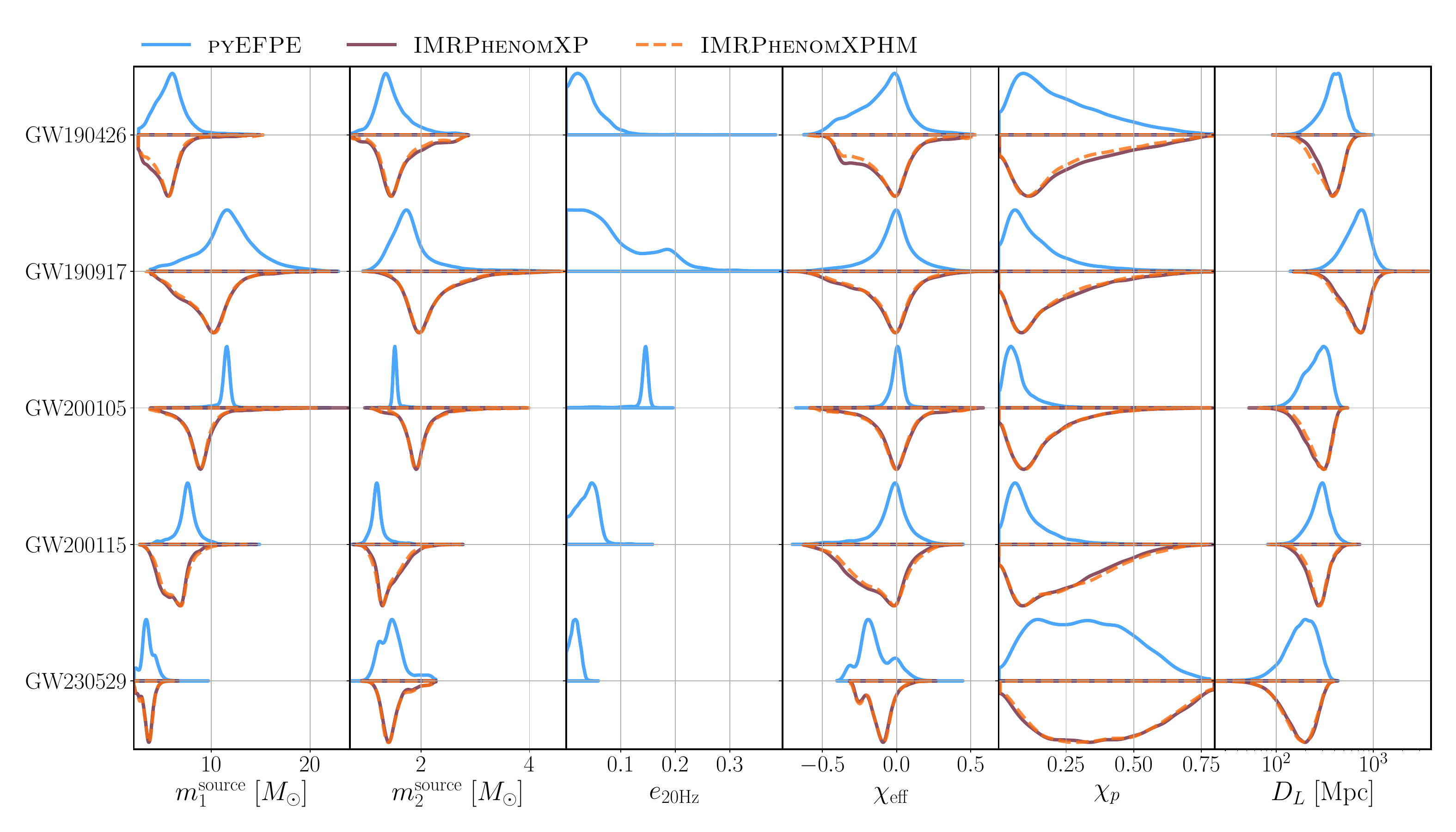}
    \caption{Marginal posterior distributions for selected source parameters of the selected NSBH events. Shown are the source-frame component masses ($m_1^\mathrm{source}$ and $m_2^\mathrm{source}$), eccentricity at 20Hz ($e_{20\mathrm{Hz}}$), the effective inspiral spin ($\chi_\mathrm{eff}$), the effective precession spin ($\chi_p$), and the luminosity distance ($D_L$). In each violin plot, the upper half corresponds to results from the eccentric \pyEFPE analysis, while the lower half shows posteriors from the quasi-circular \XP (solid) and \XPHM (dashed) analyses.}
    \label{fig:NSBHs_violins}
\end{figure*}

The marginal posterior distributions for selected source parameters of the NSBH events are shown in Fig.~\ref{fig:NSBHs_violins}. Among these, GW200105 stands out as a clear outlier, being the only event whose eccentricity posterior peaks significantly away from zero. This behavior has been studied in detail in~\citep{Morras:2025xfu,Planas:2025plq,Jan:2025fps,Kacanja:2025kpr,Tiwari:2025fua}. Notably, GW200105 also exhibits much narrower posteriors compared to the other NSBH events, likely due to the measurement of multiple orbital harmonics~\citep{Morras:2025nlp}. 

The remaining events are consistent with being quasi-circular. In particular, the eccentricity posteriors of GW200115 and GW230529 are in good agreement with those reported in~\citet{Planas:2025plq} and \citet{Kacanja:2025kpr}, while, to our knowledge, no previous eccentric analyses have been performed for GW190426 and GW190917.

Excluding GW200105, whose inferred parameters are significantly modified by the presence of eccentricity, the remaining events in Fig.~\ref{fig:NSBHs_violins} show broad agreement between the parameters estimated with \pyEFPE and those obtained using the quasi-circular \XP and \XPHM models. In general, these systems are consistent with being non-spinning, with primary masses that are relatively low compared to the black holes observed in binary black hole mergers~\citep{LIGOScientific:2025pvj}, and secondary masses that span the full $\sim 1$--$2.3\,M_\odot$ range where neutron stars are expected to reside~\citep{Kiziltan:2013oja,Landry:2021hvl}.

%---------- Astrophysical Implications
\section{Astrophysical Implications}

\begin{figure*}[th!]
    \centering
    \includegraphics[scale=0.5]{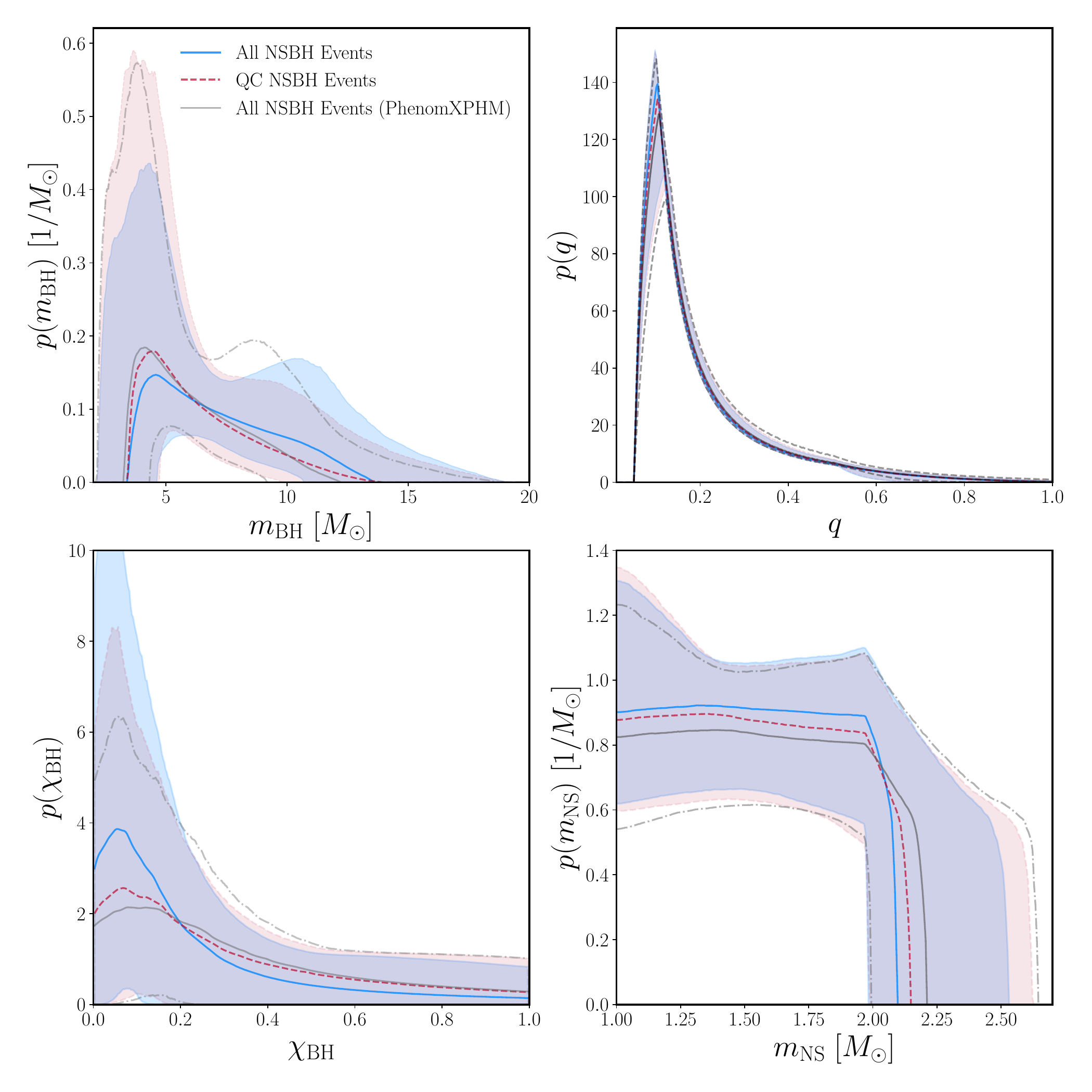}
    \caption{
        The median and 90\% credible intervals of the population predictive distributions for the black hole mass, the mass ratio,
        the black hole spin, and the neutron star mass. The blue curves analyses all NSBH events using pyEFPE, the red curve only the quasi-circular NSBH events, and the grey curve all NSBH events but using IMRPhenomXPHM (i.e. neglecting eccentricity).
    }
    \label{fig:NSBH_PPD_properties}
\end{figure*}

\subsection{Methodology}
We characterize the population properties of astrophysical NSBH systems using 
individual event source parameters inferred with the \pyEFPE model, employing 
hierarchical Bayesian inference following~\citep{Mandel:2018mve, Thrane:2018qnx, 
KAGRA:2021duu, Biscoveanu:2022iue, Callister:2024cdx}.

Given data $\setdata$ consisting of $N_{\rm det}$ detected GW events, we infer 
the posterior on hyperparameters $\Lambda$ characterizing the astrophysical population,
\begin{align}
    p(\Lambda | \setdata) = \frac{\mathcal{L}(\setdata |\Lambda) \, \pi(\Lambda)}{\mathcal{Z}_{\Lambda}},
\end{align}
where $\pi(\Lambda)$ and $\mathcal{Z}_{\Lambda}$ are the hyperparameter prior and 
evidence, respectively. Modeling the detected events as an inhomogeneous Poisson 
process~\citep{Mandel:2018mve}, the likelihood takes the form
\begin{align}
    \mathcal{L}(\lbrace d \rbrace, N_{\rm det} | \Lambda) \propto\; &N(\Lambda)^{N_{\rm det}} 
    e^{-N_{\rm exp}(\Lambda)} \nonumber \\
    &\times \prod^{N_{\rm det}}_{i=1} \int d\theta\; \mathcal{L}(d_i | \theta)\, \pi(\theta | \Lambda),
\end{align}
where $N(\Lambda)$ is the expected total number of mergers (detected and undetected) during the observing 
period, and
\begin{align}
    N_{\rm exp}(\Lambda) = N(\Lambda)\,\alpha(\Lambda)
\end{align}
is the expected number of detections. 
The detectability fraction $\alpha(\Lambda)$ 
accounts for selection effects by giving the fraction of binaries drawn from the population model that would be detectable~\citep{Loredo:2004nn, Mandel:2018mve}. 
For each event $i$, $\mathcal{L}(d_i \mid \theta)$ is the likelihood of the strain 
data given source parameters $\theta$, and $\pi(\theta \mid \Lambda)$ is the 
distribution of source parameters conditioned on the population hyperparameters $\Lambda$.

We compute $\alpha(\Lambda)$ via Monte Carlo integration following~\citep{Farr:2019abc}, 
using sensitivity estimates for NSBH binaries derived from injection campaigns 
conducted in support of the LVK analyses~\citep{Abbott:2021abc,Essick:2025bna}. We note that 
these selection effects do not yet account for orbital eccentricity, which we defer 
to future work; see~\citep{Nitz:2021mzz, Dhurkunde:2023qoe, Phukon:2024amh, 
Phukon:2025cky} for progress on incorporating eccentricity into searches, and~\citep{Singh:2025ojp, Romero-Shaw:2025otx} for efforts toward characterizing the observability of eccentricity in compact binary populations.

Our exploration of the NSBH population is largely templated on the analysis by~\citet{Biscoveanu:2022iue}. 
Here we outline the specific choices made in our analysis.
We model the black hole mass distribution as a truncated power-law, such that
\begin{align}
&\pi_{}(m_{\rm BH}|\alpha, m_{\rm BH,min}, m_{\rm BH,max}) \\ &\nonumber \quad \propto
\begin{cases}
m_{\rm BH}^{-\alpha}, & m_{\rm BH,min} \leq m_{\rm BH} \leq m_{\rm BH,max} \\
0, & \text{otherwise}
\end{cases}.
\end{align}
For the mass ratio, we use the truncated Gaussian model proposed in~\citep{Fishbach:2019bbm,Biscoveanu:2022iue} 
\begin{align}
& \pi_{}(q|m_{\rm BH}, m_{\rm NS,max}, \mu, \sigma) \\ &\nonumber \, \propto
\begin{cases}
\mathcal{N}(q|\mu, \sigma), & q_{\rm min}(m_{\rm BH}) \leq q \leq q_{\rm max}(m_{\rm BH}, m_{\rm NS,max}) \\
0, & \text{otherwise}
\end{cases},
\end{align}
which describes the conditional pairing distribution between black holes and the neutron stars. 
As in~\citep{Biscoveanu:2022iue}, we assume that the black hole is always the more massive of the two compact objects in the binary with the minimum neutron star mass taken to be $1 M_{\odot}$ such that $q_{\rm min} = 1 / m_{\rm BH, max}$.
The maximum neutron star mass is taken as a free hyperparameter to be constrained by the data. 

Following~\citep{LIGOScientific:2025pvj}, the black hole spin magnitude is taken to follow a truncated Normal distribution parameterized by $\lbrace \mu_{\chi}, \sigma_{\chi} \rbrace$,
\begin{align}
    \pi_{}(\chi_{\rm BH} | \mu_{\chi}, \sigma_{\chi} ) = \mathcal{N}_{[0,1]}(\chi_{\rm BH} | \mu_{\chi}, \sigma_{\chi}).
\end{align}
We adopt the truncated Normal distribution over the Beta distributions, see~\citep{Wysocki:2018mpo}, as these are known to enforce $p(\chi) = 0$ at $\chi = 0$ and $\chi=1$, failing to model contributions to the population at near-zero and near-extremal spins respectively~\citep{LIGOScientific:2025pvj}. 
The black hole spin orientation, denoted by $z_{\rm BH} = \cos \theta_{\rm BH}$, is modeled as an admixture of two populations~\citep{Talbot:2017yur}: a population with spins preferentially aligned with the orbital angular momentum and a population with isotropically oriented spins. 
These are intended to broadly capture contributions from the isolated evolution channel and contributions from dynamical formation scenarios.
Adopting the model proposed in~\citep{Vitale:2022dpa}, the preferentially aligned population is described by a truncated Gaussian in $z_{\rm BH}$ with mean $\mu_t$ and standard deviation $\sigma_{t}$, such that
\begin{align}
\pi(z_{\rm BH}\mid \xi,\mu_t,\sigma_t)
&=
(1-\xi)\,\frac{1}{2}
+
\xi\,\mathcal{N}_{[-1,1]}\!\left(z_{\rm BH}\mid \mu_t,\sigma_t\right).
\end{align}
Here, $\xi$ denotes the fraction of binaries in the preferentially aligned population.
We do not model the neutron star spin distribution, as we expect this to remain largely unresolved given current detector sensitivities. 
Instead, following~\citep{Biscoveanu:2022iue}, we adopt a physical constraint that the neutron star spins remains below the mass-shedding limit of $\chi_{\rm Kep} \sim 0.7$ \citep{Shao:2019ioq, Most:2020bba}.

For the eccentricity, in lieu of a more astrophysically motivated model, we explore two different population models. The first model, adopted as our default model, introduces a truncated normal to bound the mean and variance of the eccentricity in the population
\begin{align}
\label{eq:default_ecc}
& \pi_{}(e|\mu_{e}, \sigma_{e}) \propto
\begin{cases}
\mathcal{N}(e|\mu_{e}, \sigma_{e}), & 0 \leq e \leq 1 \\
0, & \text{otherwise}
\end{cases}.
\end{align}
We emphasize that the eccentricity distribution adopted here is a phenomenological parametrization rather than an astrophysically motivated model. 
Given the paucity of confidently measured eccentric binaries, the aim is to quantify how far from zero the population is allowed to deviate, without imposing any specific physical form on the underlying distribution.

A related model is the generalization to an admixture model designed to capture contributions from low-eccentricity (isolated evolution) and high-eccentricity (dynamical) formation channels,
\begin{align}
\label{eq:mixture}
&\pi(e \mid \lambda_{\rm qc}, \mu_{\rm dyn}, \sigma_{\rm qc}, \sigma_{\rm dyn})
=
\lambda_{\rm qc}\,
\mathcal{N}_{[0,1]}\!\left(e \mid 0, \sigma_{\rm qc}\right)
\nonumber\\
&\qquad
+ \;
(1-\lambda_{\rm qc})\,
\mathcal{N}_{[0,1]}\!\left(e \mid \mu_{\rm dyn}, \sigma_{\rm dyn}\right).
\end{align}
The quasi-circular population is parameterized by $\lbrace \mu_{\rm qc} = 0, \sigma_{\rm qc} \rbrace$, and the dynamical population by $\lbrace \mu_{\rm dyn}, \sigma_{\rm dyn} \rbrace$. The fraction of binaries in the quasi-circular population is controlled by the hyperparameter $\lambda_{\rm qc}$.

In practice, we use the \textsc{GWPopulation} package~\citep{Talbot:2019abc} together with the nested sampler \textsc{dynesty}~\citep{Speagle:2020abc}, as implemented in \textsc{Bilby}~\citep{Ashton:2018jfp}, to draw posterior samples from the distribution $p(\Lambda | \setdata)$.
We use uniform priors on all population hyperparameters, with the ranges reported in Table~\ref{tab:hyperpriors}.
Population predictive distributions (PPD),
\begin{align}
p(\theta | \setdata) &= \int \pi(\theta | \Lambda) p(\Lambda | \setdata) d\Lambda,
\end{align}
are used to show the impact of assumptions on event selection on the anticipated astrophysical distribution of binary source parameters $\theta$ conditional on the observed events. 

\subsection{Population Properties}
Before discussing the inferred astrophysical population properties, we emphasize that current NSBH population inferences rely on only a handful of confidently identified events. As a consequence, the inferred distributions of masses, spins, and eccentricities remain sensitive to individual systems, and population-level conclusions should be interpreted with appropriate caution. At the same time, the presence of even a single confidently measured eccentric NSBH merger highlights the potential importance of dynamical formation pathways and motivates comparison with theoretical predictions. As the number of NSBH detections grows, our ability to disentangle the contributions from multiple channels will improve substantially.

Fundamentally, we expect a broad range of astrophysical formation channels to contribute to the overall NSBH merger rate, each of which will predict a range of masses, spins, and eccentricities.
With only a limited set of NSBH observations, we cannot meaningfully extract the branching ratios. 
To interpret the observed NSBH population, we therefore consider a few broad hypotheses. In the first, we assume that all NSBH mergers originate from a single underlying population and therefore analyse the entire sample together. Although the significant eccentricity of GW200105 rules out an origin through isolated binary evolution, the properties of the full observed NSBH sample remain consistent with expectations from hierarchical triple evolution. In particular, \citet{Stegmann:2025clo} show that hierarchical triples can reproduce the observed NSBH merger rate and key population features, including large spin-orbit misalignment and residual eccentricity, with the probability of generating an event with $e_{20} \geq 0.145$, as inferred for GW200105~\citep{Morras:2025xfu}, being approximately $25\%$. 
In this framework, all NSBH events observed to date, including the eccentric binary, can be explained without invoking additional formation channels. 
This allows us to treat the full catalogue as a single population in this scenario, while noting that a more detailed accounting of channel contributions will require more detections.

In the second hypothesis, we assume that the eccentric event GW200105 represents a distinct formation pathway and therefore exclude it from the population analysis. 
The remaining NSBH mergers exhibit properties consistent with those expected from isolated binary evolution~\citep{Broekgaarden:2021iew,Broekgaarden:2021hlu}. 
We refer to this subset as the quasi-circular (QC) population and use it to characterize the mass distribution, spin orientations, and expected residual eccentricities of NSBH binaries that plausibly arise from isolated evolution. 
The QC population is anticipated to exhibit low eccentricity, efficient tidal circularization, and preferential spin-orbit alignment due to the binary's stable evolutionary history. 
By isolating this subset, we can assess whether the observed QC events are consistent with standard isolated formation channels, and explore the extent to which their inferred population properties differ from those obtained when all events are treated as part of a single, coherent NSBH population. 
Within this quasi-circular subset, we find no compelling deviations from the expectations of isolated binary evolution: the inferred eccentricities remain consistent with efficient tidal circularisation, the spins show no evidence for strong misalignment, and the overall source properties broadly match those predicted by standard field-binary formation models~\citep{Broekgaarden:2021iew}. 

The final scenario we consider employs a mixture model for the eccentricity distribution, as given in Eq.~\ref{eq:mixture}. 
In this scenario, we model two distinct contributions to the NSBH population arising from low-eccentricity (isolated binary evolution) and high-eccentricity (dynamical) formation channels. 
Whilst the eccentricity distribution is modeled by two components, we do not attempt a channel-by-channel decomposition of the mass and spin distributions. 
This is a pragmatic choice, as the eccentricity of GW200105 provides a clear discriminating observable that motivates the two-component model. 
In contrast, the mass and spin distributions predicted by different formation channels are expected to be highly degenerate.
Given the limited number of current NSBH detections, the data lack the statistical power to meaningfully separate channel contributions in these parameters, and we therefore characterize the mass and spin properties of the \textit{total} NSBH population without distinguishing between formation channels.

Having outlined the population scenarios under consideration, we now explore the inferred population properties, beginning with the black hole mass distribution. 
Unless otherwise stated, our comparisons will be between the first two scenarios discussed above, with the third scenario only entering into our discussion on eccentricity. 
As GW230529 is the lowest-mass NSBH observed to date, including this event shifts the minimum black hole mass to lower values, finding $m^{\rm BH}_{\rm min} = 3.57^{+1.17}_{-1.38}\,M_{\odot}$ for the total NSBH population and $3.58^{+1.21}_{-1.38}\,M_{\odot}$ for the QC population. 
This is consistent with the lower black hole mass inferred in the GW230529 discovery paper~\citep{LIGOScientific:2024elc}.  
For the upper black hole mass, whether or not we include GW200105 only leads to a relatively small impact on the overall population.
In ~\citep{Morras:2025xfu}, the mass of the black hole was reported to be $m_{\rm BH} = \massprimary$, driving a peak in the population predictive distribution just above $10 M_{\odot}$.  
Including GW200105 leads to a slight excess in more massive black hole masses and a slightly tighter bound on the maximum black hole mass of $m^{\rm BH}_{\rm max} = 13.66^{+5.48}_{-3.20} M_{\odot}$ compared to $m^{\rm BH}_{\rm max} = 13.95^{+5.29}_{-4.27} M_{\odot}$ when excluding the event. 
The shift towards a heavier upper limit in the black hole mass distribution relative to~\citep{Biscoveanu:2022iue} can be partially attributed to the shift to higher primary masses when using \pyEFPE, as seen in Fig.~\ref{fig:NSBHs_violins}.

The mass-ratio distribution is strongly driven by the priors on the maximum and minimum component masses, as discussed in~\citep{Biscoveanu:2022iue}. 
Whilst the black hole mass and spin distributions are comparatively well constrained, the prior-dependence in the mass-ratio impacts the neutron star mass distribution in a non-trivial way.
Adopting the priors used in~\citep{Biscoveanu:2022iue}, the black holes are allowed to span a range $m_{\rm BH} \in [2, 20] M_{\odot}$ and neutron stars $m_{\rm NS} \in [1, 2.7] M_{\odot}$. resulting in the mass-ratio prior clustering around $q \sim 0.1$.
We only consider the Gaussian model, as even with the additional events considered here, little information is gained relative to the earlier analysis in~\citep{Biscoveanu:2022iue}. 
We find constraints on the maximum NS mass to be $m^{\rm NS}_{\rm max} = 2.10^{+0.46}_{-0.12}$ when including GW200105 and $m^{\rm NS}_{\rm max} = 2.15^{+0.47}_{-0.16}$ when excluding the event. 
We caveat that this is the maximum NS mass as inferred from the NSBH population in isolation, and does not take into account the BNS observations, notably GW190425 which has support for NS masses up to $m^{\rm max}_{\rm NS} \approx 2.52 M_{\odot}$ assuming a high-spin prior~\citep{LIGOScientific:2020aai}.

The distribution of the black hole spin magnitudes was modeled using a truncated Gaussian rather than a Beta distribution, which can help capture contributions to the population near $\chi \approx 0$~\citep{LIGOScientific:2025pvj}.
Consistent with earlier analyses, we find that the BH spin magnitudes in NSBH are systematically smaller than those inferred from the binary black hole (BBH) population~\citep{LIGOScientific:2025pvj}. 
Recent updated analyses of GW200105, see~\citep{Morras:2025xfu}, favour small black hole spins, finding $\chi_{\rm BH} = \spinprimary$ or, in terms of the effective spin parameters, $\chi_{\rm eff} = \chieffec$ and $\chi_{p} = \chiprec$.
Due to the extremely well-constrained low-spin measurement, whether or not we include GW200105 in the population has a notable impact on the $\chi_{\rm BH}$ distribution. 
Nevertheless, the population predictive distribution for the black‑hole spin magnitudes continues to favour low $\chi_{\rm BH}$, consistent with the emerging picture that NSBH systems may host systematically lower spins than those observed in BBH mergers.

\begin{figure}
    \centering
    \includegraphics[width=1\columnwidth]{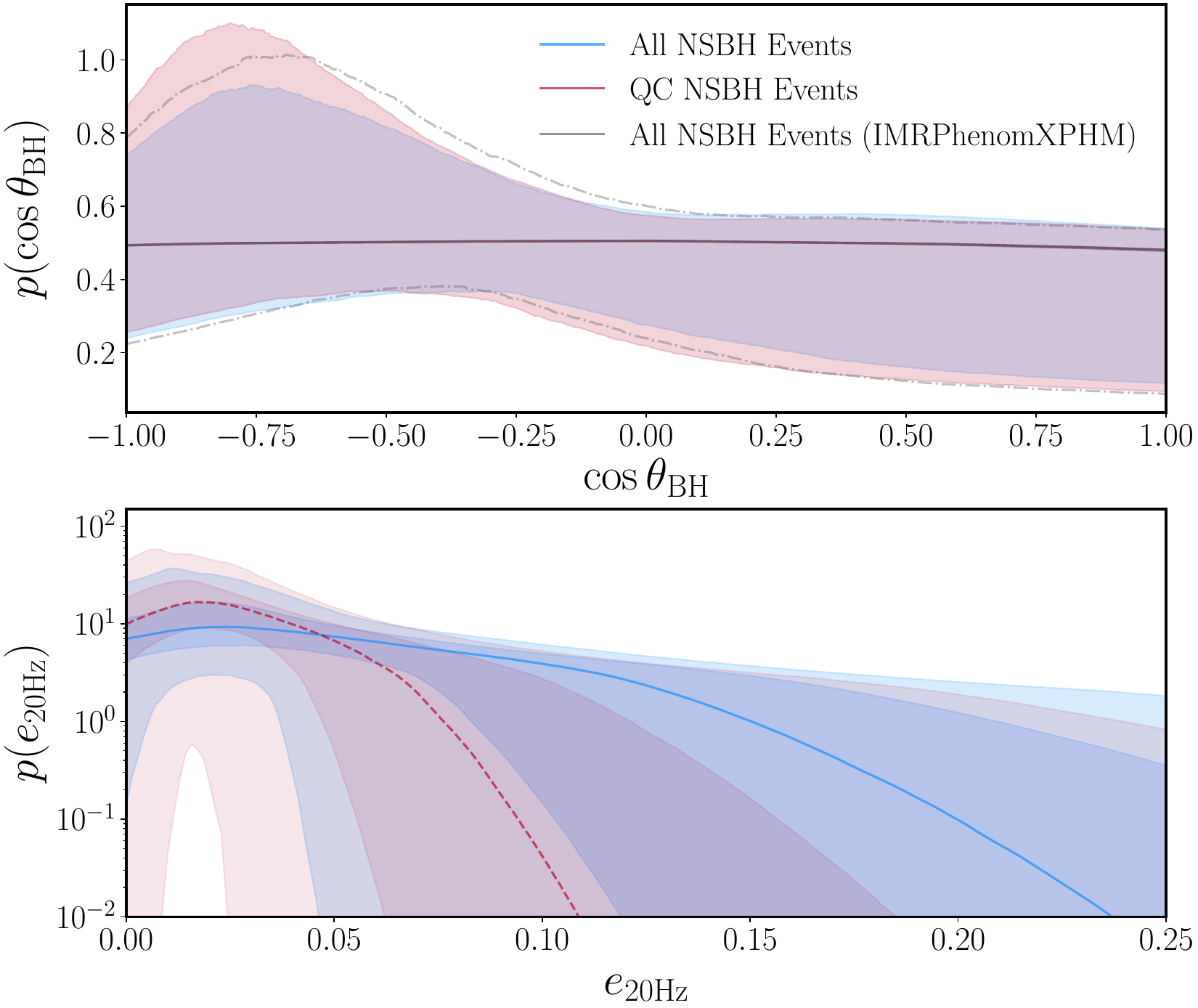}
    \caption{Population predictive distributions for the truncated Normal population model for eccentricity. The top panel shows the cosine of the black hole spin tilt angle $\cos \theta_{\rm BH}$ and the lower panel  the eccentricity at $20$Hz.
    }
    \label{fig:NSBH_PPD_ecc_so}
\end{figure}

For the spin-tilt distribution, unlike the BBH observations~\citep{LIGOScientific:2025pvj,Stegmann:2025zkb}, the current NSBH population is only weakly constrained. 
Interestingly, even after excluding the GW200105, the quasi‑circular NSBH binaries exhibit a mild preference for $\cos\theta_{\rm BH} < 0$, see the bottom panel of Fig.~\ref{fig:NSBH_PPD_ecc_so}, indicating a tendency toward spin–orbit misalignment. 
Although this trend is not yet statistically significant, it is qualitatively consistent with expectations from dynamically influenced formation channels.
However, there are several factors that must be taken into account. 
For low-SNR events with $\chi_{\rm eff} \approx 0$, and poorly constrained $\chi_p$, the likelihood provides limited information about spin orientations. 
As such, the event‑level posteriors are largely shaped by the prior volume, and the hierarchical inference for these parameters remains prior‑dominated. 
This is coupled with a known degeneracy between the masses and spins of the binary components~\citep{Cutler:1994ys,LIGOScientific:2024elc}, in which more comparable mass ratios correlate with a more negative $\chi_{\rm eff}$, though the inclusion of spin-precession can help break this degeneracy~\citep{Vecchio:2003tn,Chatziioannou:2014coa,Pratten:2020igi}.
This effect can be partially seen in Fig.~\ref{fig:NSBHs_violins}, where the posterior distributions for $\chi_{\rm eff}$ show a tail that is mildly negatively skewed, see also the discussion in~\citep{LIGOScientific:2024elc}.
As a result, the mild preference for negative $\cos\theta_{\rm BH}$ at the population level could be interpreted as a consequence of these prior‑driven and degeneracy‑driven effects, rather than as evidence for astrophysical spin-orbit misalignment, warranting further study.

In the hierarchical triple analysis presented in~\citet{Stegmann:2025clo}, tertiary-induced NSBH mergers were found to occur in a strongly non-adiabatic regime in which the ZKL timescale is much shorter than the relativistic spin-precession timescale, i.e., $\tau_{\rm ZKL} \ll \tau_{p}$. 
In this limit, the black hole spin precesses about the total angular momentum $\mathbf{J}$ of the system while maintaining an approximately constant tilt angle relative to $\mathbf{J}$, whereas the orbital angular momentum vector undergoes large‑amplitude reorientation.
This naturally leads to a $\cos\theta_{\rm BH}$ distribution that is broad and approximately symmetric about zero, leading to a non-negligible fraction of mergers ($\sim\!10$--$20\%$ in the triple-channel models) with retrograde spin configurations.
This behaviour is consistent with recent analyses of non-adiabatic spin dynamics, such as~\citet{Rodriguez:2018jqu}, who showed that when the spin–precession timescale exceeds the ZKL timescale, the spin remains effectively fixed while the orbital plane reorients, naturally producing broad spin–tilt distributions.
The population predictive distributions in Fig.~\ref{fig:NSBH_PPD_ecc_so} are in broad agreement with these predictions.

By contrast, isolated binary evolution through a common-envelope phase is expected to produce a distribution sharply peaked near $\cos\theta_{\rm BH} \simeq 1$, unless large natal kicks are invoked. 
In particular,~\citep{Fragione:2021qtg} argued that for common-envelope evolution to result in large spin-orbit tilts, one would need natal kicks $\gtrsim 150 \, \rm{km} \, \rm{s}^{-1}$ and highly efficient common-envelope ejection, $\alpha_{\rm CE} \gtrsim 3$.
Whilst our results are fairly uninformative, there is a slight tension with the isolated evolution predictions due to the mild preference for $\cos \theta_{\rm BH} < 0$, as discussed above.

Dynamical assembly in star clusters is also expected to produce broadly distributed, and in some cases nearly isotropic, spin–tilt orientations \citep[e.g.][]{Rodriguez:2016vmx,Hoang:2020gsi}, but the NSBH merger rate from clusters is expected to be orders of magnitude below the inferred rates~\citep{LIGOScientific:2025pvj}, making this channel an unlikely contributor to the observed population.

% Now to eccentricity
For the eccentricity distributions, GW200105 shifts the mean of the truncated Normal model toward larger eccentricities, generating an extended high-eccentricity tail in the posterior predictive distribution (PPD; bottom panel of Fig.~\ref{fig:NSBH_PPD_ecc_so}). 
We note that the $90$\% credible intervals are sensitive to the lower prior bound on the variance of the truncated Normal, $\sigma_{e}$, leading to the apparent gap in the predictive distribution around $e_{20\rm{Hz}} \sim 0.02$.
After marginalizing over hyperparameter uncertainty, we place $90$\% upper limits on the population eccentricity at 20,Hz of
$e_{\rm 20Hz} < 0.11^{+0.17}{-0.09}$ for the full NSBH sample and
$e_{\rm 20Hz} < 0.05^{+0.15}_{-0.04}$ for the quasi-circular (QC) subset.
The quoted uncertainties correspond to the $90$\% credible interval of the posterior predictive distributions.

\begin{figure}
    \centering
    \includegraphics[width=1\columnwidth]{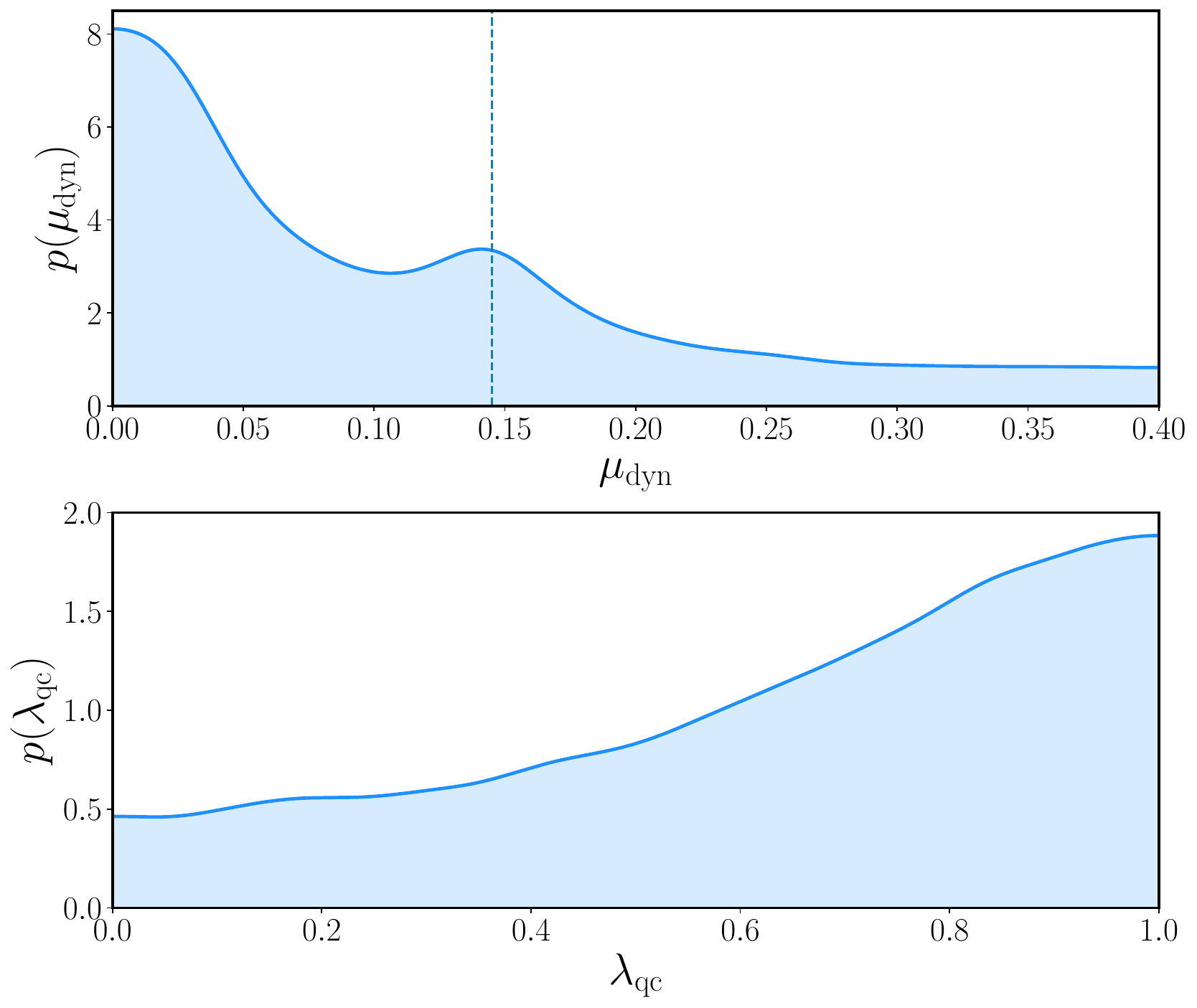}
    \caption{Select hyperposteriors for the truncated Normal mixture model, representing an eccentric (dynamical) channel and a quasi-circular (isolated) channel. The upper panel shows the population mean eccentricity of the dynamical component $\mu_{\rm dyn}$, exhibiting a mild peak near the value inferred for GW200105 \citep{Morras:2025xfu}, denoted by the vertical dashed line. The lower panel displays the mixing fraction $\lambda_{\rm qc}$ between the two channels. This parameter is only weakly constrained, but does disfavor a purely dynamical population, consistent with event-level parameter estimation.
    }
    \label{fig:NSBH_PPD_ecc_mix}
\end{figure}

Figure~\ref{fig:NSBH_PPD_ecc_mix} presents a subset of the hyperparameters from the mixture model described in Eq.~\ref{eq:mixture}. 
The upper panel shows the population mean eccentricity of the dynamical component, $\mu_{\rm dyn}$, which exhibits a mild peak near the value inferred eccentricity of GW200105~\citep{Morras:2025xfu}. 
The lower panel displays the mixing fraction between the quasi-circular and dynamical channels, $\lambda_{\rm qc}$. 
Although this parameter remains weakly constrained, the posterior disfavors a purely dynamical population, consistent with the conclusions from event-level parameter estimation.
The variances of the truncated Normal components are likewise only weakly constrained, as expected given the limited number of observations. 
The remaining population hyperparameters are consistent with those obtained under the default truncated Normal model (Eq.~\ref{eq:default_ecc}), which is again expected given that we do not model correlations among these parameters.

%---------- Distributions

\begin{figure}
    \centering
    \includegraphics[width=1\columnwidth]{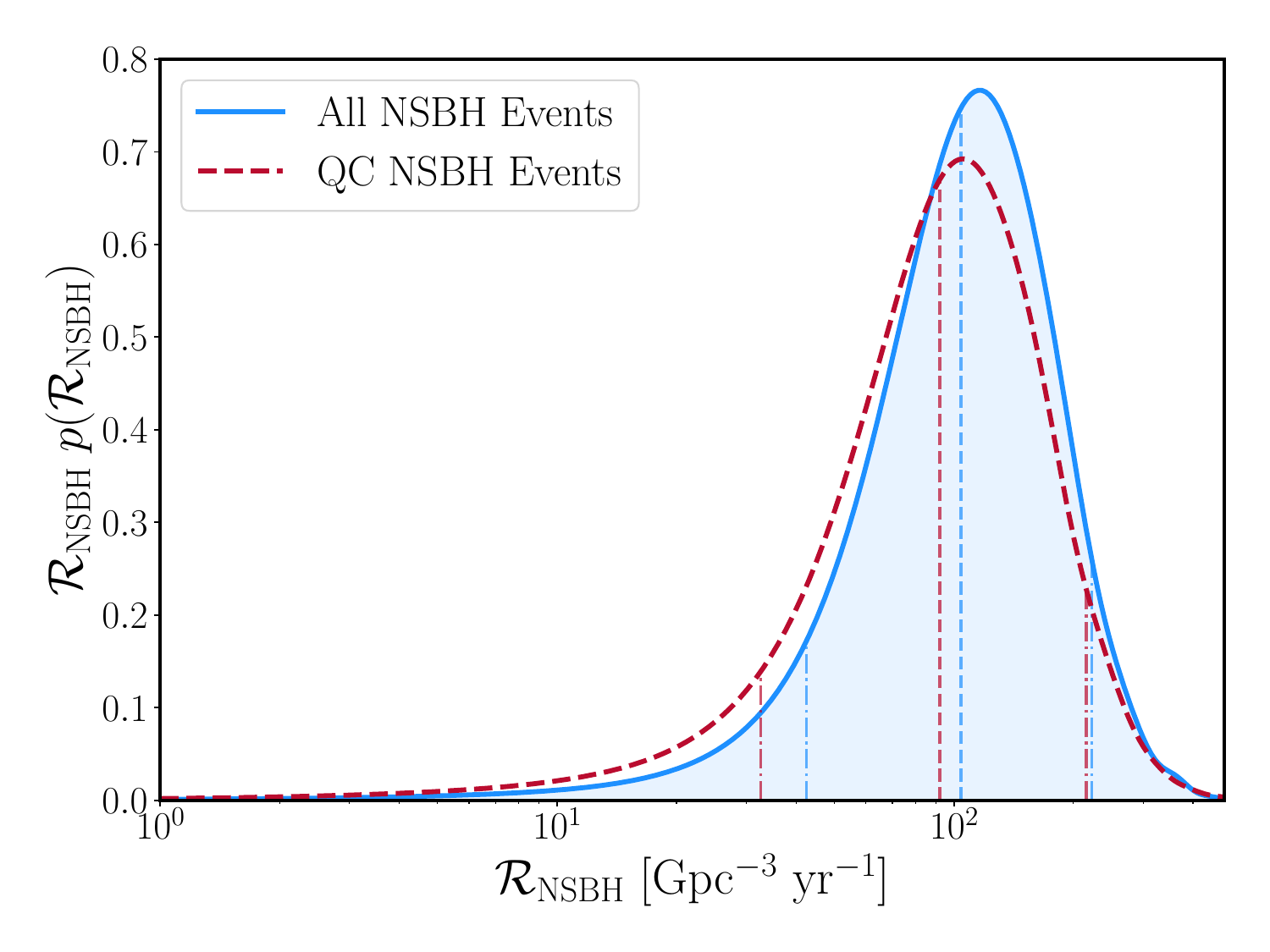}
    \caption{Merger rates for the NSBH population under two assumptions. First, that GW200105 is treated as part of a common NSBH population (blue). 
    Second, that the orbital eccentricity observed in GW200105 suggests that if formed through a dynamical channel and should be treated separately and is excluded from the analysis (red, dashed). The vertical lines denote the median and $90$\% credible interval.
    }
    \label{fig:NSBH_rates}
\end{figure}

%---------- Rates
Assuming a scale-invariant prior on the merger rate, $\pi(\mathcal{R}_{\rm NSBH}) \propto 1/\mathcal{R}_{\rm NSBH}$, 
we infer an overall neutron star–black hole merger rate of 
$\mathcal{R}_{\rm NSBH} = \rateALNSBH~\mathrm{Gpc}^{-3}\,\mathrm{yr}^{-1}$, 
and $\mathcal{R}_{\rm NSBH} = \rateQCNSBH~\mathrm{Gpc}^{-3}\,\mathrm{yr}^{-1}$ 
when restricting the analysis to quasi-circular binaries, as shown in 
Fig.~\ref{fig:NSBH_rates}. 
Including or excluding the eccentric event GW200105 has only a minimal impact on the inferred merger rate.
These rates are broadly consistent with those reported in~\citep{LIGOScientific:2024elc,LIGOScientific:2025pvj}, though we emphasise that these constraints remain limited by the small 
number of confidently identified NSBH mergers, and caution that individual events can disproportionately influence population-level inferences. 

% Closing caveats
As a final caveat, our population model does not incorporate the full set of correlations that may exist among the hyperparameters. 
For simplicity, we assume that the masses, spins, and orbital eccentricities are independent at the population level, reducing the dimensionality of the hyperparameter space. 
However, most astrophysical formation scenarios naturally predict correlated structure in these parameters. 
For example, correlations between the mass ratio $q$ and the effective inspiral spin $\chi_{\rm eff}$ have been identified in the binary black hole population~\citep{Callister:2021fpo,Adamcewicz:2022hce,LIGOScientific:2025pvj} and there is evidence that the spin distribution broadens with redshift~\citep{Biscoveanu:2022qac}.
Similar correlations involving eccentricity, spin-orbit misalignment, and component masses are expected, given our understanding of dynamical or triple-induced formation pathways~\citep{Stegmann:2025clo}.

Neglecting such correlations can influence hierarchical inference in several ways, due to model misspecification~\citep{Payne:2022xan}. 
Correlated subpopulations projected onto independent, marginalized one-dimensional distributions may artificially broaden the posteriors, while correlated features may mimic or obscure signatures associated with a particular formation channels. 
In turn, this can bias our inference on the branching ratios or mixing fractions. 
With only a handful of NSBH systems having been observed, the data cannot yet meaningfully constrain higher-dimensional correlated population models, and any inferred correlations would be poorly determined.
Nevertheless, as the number of NSBH observations grows, incorporating correlated structures will become essential for accurate astrophysical inference.

%---------- Conclusions
\section{Conclusions}
We revisited the population properties of neutron star–black hole mergers using eccentric, precessing waveform analyses of all currently reported low-mass compact binary coalescences. Under a scale-invariant prior on the merger rate, we inferred consistent rates whether analysing the full sample or restricting to the quasi-circular subset, indicating that the overall rate estimate is not driven by the inclusion of the eccentric event GW200105.

Treating all NSBH mergers as a single population remains compatible with dynamically influenced formation channels. 
In particular, hierarchical triples naturally reproduce several qualitative features of the observed population~\citep{Stegmann:2025clo}, including measurable residual eccentricity and broad spin–orbit misalignment without requiring fine tuning.

At the same time, the quasi-circular subset shows no compelling deviations from expectations based on isolated binary evolution. 
Whilst there is a mild preference for $\cos \theta_{\rm BH} < 0$, which would be in tension with the predictions from isolated evolution, we do not yet have enough information to make robust astrophysical statements or to decouple this from prior- and correlation-driven effects.
The inferred eccentricities are consistent with efficient circularisation and the component masses fall within ranges compatible with standard field-binary formation. 
These systems may therefore represent a subpopulation consistent with canonical isolated evolution~\citep{Broekgaarden:2021iew}.

The eccentricity observed in GW200105, however, cannot be reconciled with isolated evolution and must be accommodated for in any complete formation picture. 
This suggests either a single formation pathway capable of producing both quasi-circular and eccentric systems, or the presence of multiple subpopulations with distinct evolutionary origins. 
The current sample is too small to robustly discriminate between these possibilities, but it is already clear that field-binary evolution alone cannot account for the full diversity of observed NSBH mergers.

As of now, population-level conclusions on NSBH binaries remains limited by small-number statistics. 
Although GW230529 lowers the minimum black hole mass and GW200105 influences the inferred spin-magnitude distribution, the overall sample still favours comparatively low black hole spins and only weakly constrains spin–orbit orientations. 
Additional observations will be required to determine whether these trends persist, and to quantify the relative contributions of isolated and dynamical channels.

We have entered an exciting period of gravitational‑wave astronomy, in which orbital eccentricity is starting to become a uniquely powerful tracer of compact‑binary assembly. 
Unlike masses or spins, which can be degenerate across binary formation channels, measurable eccentricity in the LIGO–Virgo–KAGRA band offers a direct signature of dynamical formation in environments such as globular clusters, nuclear star clusters, hierarchical triples, or AGN disks, while isolated evolution predicts near‑circular orbits by the time systems are detectable. 
As sensitivity improves, and the sample of events grows, joint population inference on rates, masses, spins, and eccentricity will enable far more definitive discrimination between these pathways, especially when correlations between these parameters are taken into account. 
This progress positions NSBH binaries as particularly valuable probes of the astrophysical conditions that shape compact binary mergers.

%---------- Acknowledgements
\begin{acknowledgments}
The authors thank Reed Essick and Ivan Markin for useful comments on the manuscript. 
G.M. acknowledges support from the Ministerio de Universidades through Grants No. FPU20/02857 and No. EST24/00099, and from the Agencia Estatal de Investigaci\'on through the Grant IFT Centro de Excelencia Severo Ochoa No. CEX2020-001007-S, funded by MCIN/AEI/10.13039/501100011033.
G.P. is very grateful for support from a Royal Society University Research Fellowship URF{\textbackslash}R1{\textbackslash}221500 and RF{\textbackslash}ERE{\textbackslash}221015, and gratefully acknowledges support from an NVIDIA Academic Hardware Grant.
G.P. and P.S. acknowledge support from STFC grants ST/V005677/1 and ST/Y00423X/1, and a UK Space Agency grant ST/Y004922/1. 
P.S. also acknowledges support from a Royal Society Research Grant RG{\textbackslash}R1{\textbackslash}241327.
The authors are grateful to the Rates and Populations group in the LVK for preliminary feedback on the research.
%CIT acknowledgements
The authors are grateful for computational resources provided by the LIGO Laboratory (CIT) and supported by the National Science Foundation Grants PHY-0757058 and PHY-0823459, the University of Birmingham's BlueBEAR HPC service, which provides a High Performance Computing service to the University's research community, as well as resources provided by Supercomputing Wales, funded by STFC grants ST/I006285/1 and ST/V001167/1 supporting the UK Involvement in the Operation of Advanced LIGO.
% NSF & GWOSC acknowledgment
This research has made use of data or software obtained from the Gravitational Wave Open Science Center (gwosc.org), a service of the LIGO Scientific Collaboration, the Virgo Collaboration, and KAGRA. This material is based upon work supported by NSF's LIGO Laboratory which is a major facility fully funded by the National Science Foundation, as well as the Science and Technology Facilities Council (STFC) of the United Kingdom, the Max-Planck-Society (MPS), and the State of Niedersachsen/Germany for support of the construction of Advanced LIGO and construction and operation of the GEO600 detector. Additional support for Advanced LIGO was provided by the Australian Research Council. Virgo is funded, through the European Gravitational Observatory (EGO), by the French Centre National de Recherche Scientifique (CNRS), the Italian Istituto Nazionale di Fisica Nucleare (INFN) and the Dutch Nikhef, with contributions by institutions from Belgium, Germany, Greece, Hungary, Ireland, Japan, Monaco, Poland, Portugal, Spain. KAGRA is supported by Ministry of Education, Culture, Sports, Science and Technology (MEXT), Japan Society for the Promotion of Science (JSPS) in Japan; National Research Foundation (NRF) and Ministry of Science and ICT (MSIT) in Korea; Academia Sinica (AS) and National Science and Technology Council (NSTC) in Taiwan.
% Code acknowledgments
Figures were prepared using \texttt{Matplotlib}~\citep{matplotlib:2007}. Analyses were performed using Bilby~\citep{bilby_paper}, Dynesty~\citep{dynesty}, LALSuite~\citep{lalsuite}, NumPy~\citep{numpy:2020}, pyEFPE~\citep{Morras:2025nlp}, and Scipy~\citep{scipy:2020}.

\end{acknowledgments}

\clearpage
\appendix
\section{Priors on Population Hyperparameters}

\begin{deluxetable}{l l c c}[th!]
\tablecaption{Prior ranges on the population hyperparameters describing the mass, spin, and eccentricity distributions used in our hierarchical Bayesian inference. All priors are taken to be uniform.
\label{tab:hyperpriors}
}
\tablehead{
\colhead{Symbol} & \colhead{Parameter} & \colhead{Minimum} & \colhead{Maximum}
}
\startdata
$\alpha$        & black hole mass power-law index      & $-4$        & $12$ \\
$m_{\rm BH,min}$ & minimum black hole mass              & $2\,M_\odot$ & $10\,M_\odot$ \\
$m_{\rm BH,max}$ & maximum black hole mass              & $8\,M_\odot$ & $20\,M_\odot$ \\
$m_{\rm NS,max}$ & maximum neutron star mass            & $1.97\,M_\odot$ & $2.7\,M_\odot$ \\
$\mu$           & mass ratio mean                       & $0.1$       & $0.6$ \\
$\sigma$        & mass ratio standard deviation         & $0.1$       & $1$ \\
$\alpha_\chi$   & black hole spin $\alpha$              & $0.1$       & $8$ \\
$\beta_\chi$    & black hole spin $\beta$               & $0.1$       & $8$ \\
$\mu_e$    & mean of eccentricity              & $0$       & $0.4$ \\
$\sigma_e$    & standard deviation of eccentricity    & $10^{-3}$       & $0.5$ \\
$\mu_{\rm dyn}$    & mean of eccentricity for the eccentric component of the mixture model     & $10^{-3}$       & $0.5$ \\
$\sigma_{\rm dyn}$    & standard deviation of eccentricity for the eccentric component of the mixture model     & $10^{-3}$       & $0.5$ \\
$\mu_{\rm qc}$    & mean of eccentricity for the quasi-circular component of the mixture model     & $0$       & $0$ \\
$\sigma_{\rm qc}$    & standard deviation of  eccentricity for the quasi-circular component of the mixture model     & $10^{-3}$       & $0.5$ \\
$\lambda_{\rm qc}$    & Mixing fraction between quasi-circular $(\lambda_{\rm qc} = 1)$ and eccentric $(\lambda_{\rm qc} = 0)$ populations               & $0$       & $1$
\enddata
\end{deluxetable}

%---------- Bibliography
\clearpage
\bibliography{references}{}
\bibliographystyle{aasjournalv7}

\end{document}